\documentclass[%
 reprint,
 amsmath,amssymb,
 aps,showkeys
]{revtex4-1}

\usepackage{graphicx}
\usepackage{dcolumn}
\usepackage{multirow}
\usepackage[table,xcdraw]{xcolor}
\usepackage{bm}
\usepackage{amssymb}
\usepackage{amsthm}
\usepackage{amsmath}
\usepackage{adjustbox}
\usepackage{wrapfig}
\usepackage{siunitx}

\makeatletter
\renewcommand{\p@subsection}{}
\renewcommand{\p@subsubsection}{}
\makeatother

\begin{document}

\preprint{APS/123-QED}

\title{\textbf{Deep Learning based Framework for Automatic Diagnosis of Glaucoma based on analysis of Focal Notching in the Optic Nerve Head}}

\author{Sneha Dasgupta\textsuperscript{a}\altaffiliation{\textit{Corresponding Author Email ID}: sneha.dasgupta.ece18@heritageit.edu.in}, Rishav Mukherjee\textsuperscript{b}, Kaushik Dutta\textsuperscript{c}}

\author{Anindya Sen\textsuperscript{a}}

\affiliation{
\textsuperscript{b} Technical University of Munich \\
\textsuperscript{c} Washington University in St.Louis \\
\textsuperscript{a} Heritage Institute of Technology, India\\
}%




\begin{abstract}
\textbf{\abstractname-} Automatic evaluation of the retinal fundus image is emerging as one of the most important
tools for early detection and treatment of progressive eye diseases like Glaucoma. Glaucoma results to a progressive degeneration of vision and is characterized by the deformation of the shape of optic cup and the degeneration of the blood vessels resulting in the formation of a notch along the neuroretinal rim. In this paper, we propose a deep learning-based pipeline for automatic segmentation of optic disc (OD) and optic cup (OC) regions from Digital Fundus Images (DFIs), thereby extracting distinct features necessary for prediction of Glaucoma. This methodology has utilized focal notch analysis of neuroretinal rim along with cup-to-disc ratio values as classifying parameters to enhance the accuracy of Computer-aided design (CAD) systems in analyzing glaucoma. Support Vector-based Machine Learning algorithm is used for classification, which classifies DFIs as Glaucomatous or Normal based on the extracted features. The proposed pipeline was evaluated on the freely available DRISHTI-GS dataset with a resultant accuracy of 93.33\% for detecting Glaucoma from DFIs.

\keywords{Glaucoma, optic-disc segmentation, Cup segmentation, cup-to-disc ratio, Focal Notching, Retinal Fundus Image, Multi-class U-Net Network, ISNT Rule, SVM, Optic Disc (OD), Optic Cup (OC) }
\end{abstract}

\maketitle


\section{\label{sec:level1}Introduction}

Glaucoma is an irreversible and chronic eye disease caused by gradual and progressive degeneration of the optical nerve fibers, leading to the structural change of the Optic Nerve Head(ONH) and subsequently causing loss of vision.\citep{michelson2008papilla}. As glaucoma cannot be cured entirely and is asymptomatic in the early stages, early detection and treatment are necessary to slow its progression.
The analysis of Digital Fundus Image(DFI) has emerged as a preferred modality of glaucoma diagnosis due to its non-invasive and economic nature, which is suitable for large-scale glaucoma screening.
\begin{figure}[h]
\centering\includegraphics[width=0.6\linewidth]{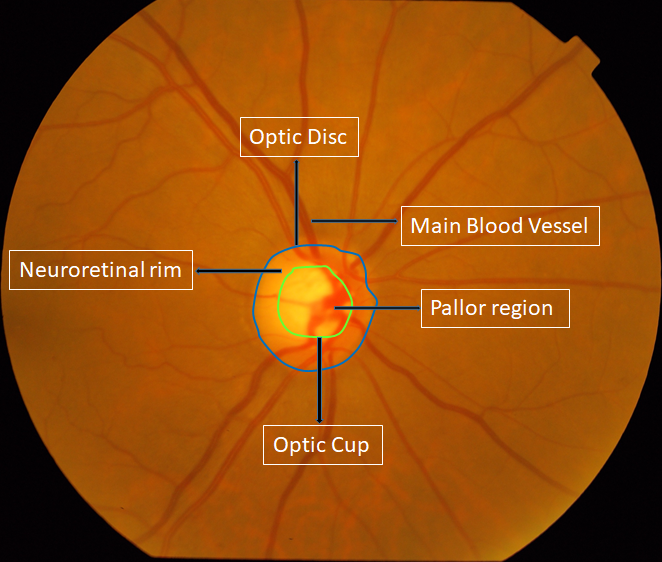}
\caption{An Optic Disc centric 2-D Retinal Image}
\end{figure}

Glaucoma is generally detected by analyzing the patient's medical history, intraocular pressure and visual field loss tests, and a manual evaluation of the Optic Disc (OD) through ophthalmoscopy. OD is a crucial component of the retina and is divided into two distinct parts, i.e., (i) the bright central depression called the cup and (ii) the peripheral region where the nerve fibers bend into the cup region called the neuroretinal rim, as shown in Figure. 1.

The loss of optic nerve fibers subsequently leads to the change
in optic disc structure, inducing enlargement of the Optic Cup (OC)
region. The process of enlarging the optic cup section and, consequently, thinning the neuroretinal rim is known as cupping. 
The enlargement of the cup region with respect to the disc diameter \citep{hancox1999optic}, peri-papillary atrophy (PPA) \citep{jonas1992glaucomatous}, the Retinal Nerve Fibre Layer (RNFL)
characteristics \citep{tuulonen1991initial}, disc \citep{Damms1993SensitivityAS},
 focal notching of the cup \citep{shields2005shields}, ISNT rule
\citep{harizman2006isnt} are considered an important
pointers for the progression of glaucoma.

Our method utilizes cup-to-disc ratio (CDR) \citep{wong2009intelligent} and neuroretinal rim width thickness (based on ISNT rule )as important classification parameters used to detect glaucoma. Usually, higher CDR values signify greater chances of glaucoma. However, the cup-to-disc ratio often fails when patients have a genetically large optic cup or myopic eye (where the optic cup is inherently large). Because of this concern, we have additionally introduced \textbf{notching} established by the ISNT rule along with the cup-to-disc ratio. Notching is a method used to measure the thickness of the neuroretinal rim.\citep{mukherjee2019predictive}\\

\begin{figure}[h]
\centering\includegraphics[width=0.5\linewidth]{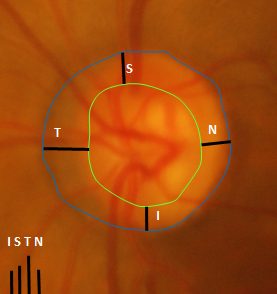}
\caption{ISNT representation for a glaucomatous left eye.}
\end{figure}

ISNT stands for the four sectors in which the optical nerve head can be segmented based on certain range of angles.
\begin{description}
  \item[$\Box$ I $\Rightarrow$ Inferior i.e. the bottom-most region]
  \item[$\Box$ S $\Rightarrow$ Superior i.e. the topmost region] 
  \item[$\Box$ N $\Rightarrow$ Nasal i.e. the near nose region] 
  \item[$\Box$ T $\Rightarrow$ Temporal i.e. the opposite of nasal region] 
\end{description}
In the case of a normal eye, the order of the thickness of the neuroretinal rim in descending order is: \[I>S>N>T\]

i.e., the inferior region has the maximum width, followed by the superior, nasal and temporal region, following the order.
In glaucoma cases, the retinal image violates this order due to abnormal elongation of the cup in inferior(I) or superior(S) region, which further results in thinning of the width of the neuroretinal rim in these two regions, as shown in figure 2.

Various researches are accounted for the localization and segmentation of OD and classification of glaucoma disorder. 
Methods based on deformable models have been introduced in \cite{lowell2004optic,osareh2002comparison,xu2007optic,wong2008level}. In \citep{lowell2004optic}, Lowell et al. applied template matching for localization of OD and a circular deformable model for segmentation. Osareh et al. \citep{osareh2002comparison} used template matching for detecting OD center approximately and thereby extracted the OD boundary using a snake initialized on a morphologically enhanced OD region. Xu et al.\citep{xu2007optic} also used a deformable model technique that includes morphological operations and an active contour model. Wong et al.\citep{wong2008level} proposed a technique that uses a modified level set method followed by ellipse fitting. 
Other approaches based on Circular Hough Transform and pixel classifications to segment the OD were proposed in \cite{abramoff2007automated,aquino2010detecting,muramatsu2011automated,dutta2018automatic}. Optic cup segmentation methods have also been introduced in \citep{xu2007optic, dutta2018automatic, wong2008level} by a method of thresholding.
However, optic cup segmentation is more challenging than disc segmentation because of the high density of blood vessels in the optic cup and disc region boundary that has further reduced the visibility of the boundary. As a result, very few methods have been proposed for cup segmentation as compared to disc segmentation.

In this paper, as shown in Figure 3, we propose a deep learning based framework for automatic segmentation of Optic Disc (OD) and Optic Cup (OC), thereby capturing the distinct features that better characterize the symptoms related to glaucoma. The segmentation of Optic Disc and Cup is the primary step in extracting different parameters from retinal fundus images necessary for the detection of glaucoma. The cup-to-disc ratio and notching characteristics(based on ISNT rule) are the learning parameters incorporated with machine learning algorithm to classify DFIs for the prediction of glaucoma.
The paper is organized as follows. In Section I, we have given an introduction to the background and inspiration for the method. In Section II, we introduce our proposed method of segmentation of OD followed by extraction of different optical image parameters and the classification method based on them. Section III shows experimental results, followed by Section IV that presents discussions and conclusions of our work.
\begin{figure*}[t!]
\centering{\includegraphics[width = \textwidth, height=7.5cm]{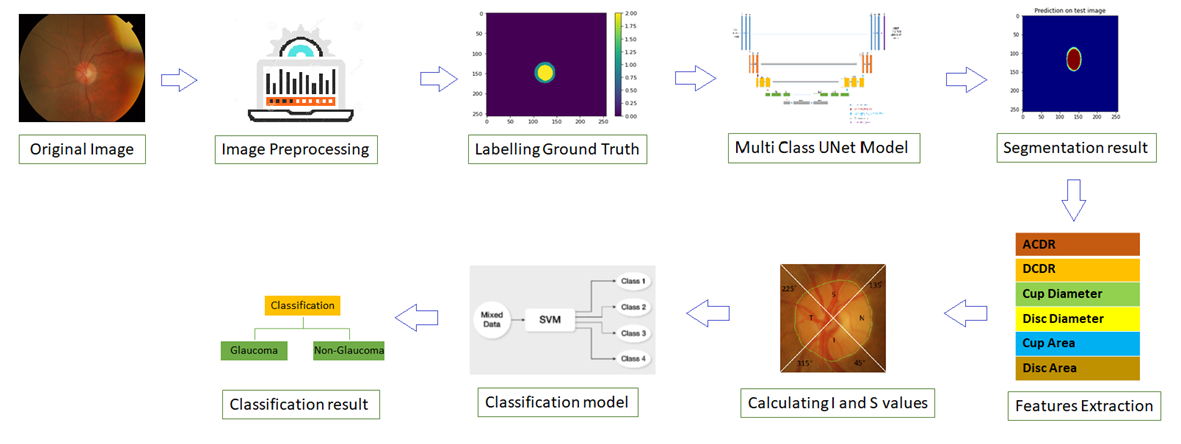}}
\caption{Deep Learning based Framework for Automatic Glaucoma Detection}
\end{figure*}

\section{\label{sec:level2}Materials and Methods}

\subsection{Dataset}
The DRISHTI-GS dataset consists of 101 fundus images with a resolution of  2896 × 1944 pixels. It comprises both Normal and Glaucomatous images along with their ground truths. This dataset has been collected and annotated by Aravind Eye Hospital, Madurai, India, in collaboration with researchers of IIIT Hyderabad. \cite{sivaswamy2015comprehensive}\cite{sivaswamy2014drishti}
In our proposed method, 71 training images of the DRISHTI-GS dataset are utilized for training the proposed model, and the rest 30 testing images are used for evaluating the performance of the final trained model.
\subsection{Image processing and data augmentation}

Since the dataset used for network training has fewer images, it may lead to overfitting. To prevent this issue, we used data augmentation to expand training images. The images used for the training proposed model are increased to 200 using data augmentation methods like horizontal flip, vertical flip, and noise addition.  About 90\% of the data-augmented training images are randomly selected to train the proposed model, and the rest 10\% images are employed for model evaluation when training the model. Furthermore, retinal images with uneven illumination and low contrast are not useful for accurate segmentation and detection of glaucoma. To overcome this challenge, we used CLAHE (Contrast Limited Adaptive Histogram Equalization) \cite{reza2004realization} as an image processing to enhance the quality of retinal images.

\subsection{Labelling Ground Truth}

At this stage, we combined the ground truth of the optic cup and optic disc provided in the DRISHTI-GS dataset to form a single mask for the fundus image as shown in Figure 4. We labeled the background of the mask as class 0, the region of the optic disc as class 1, and the optic cup region as class 2.
\begin{figure}[h]
\centering\includegraphics[width=1.0\linewidth]{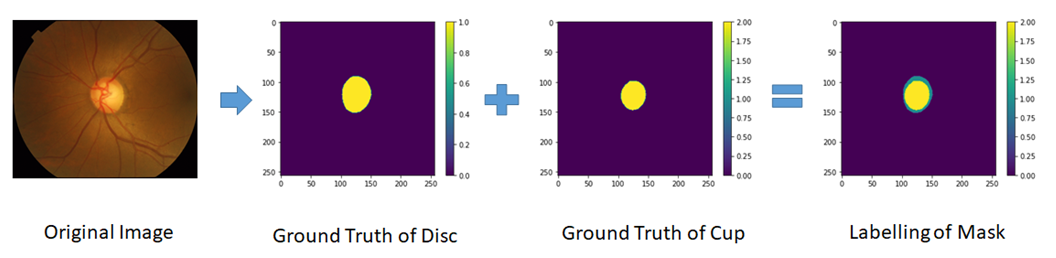}
\caption{(a)Original image (b) Binary Disc Mask (c) Binary Cup Mask (d) Multi-class OD and OC Mask}
\end{figure}

\subsection{Optic Cup Segmentation}

\begin{figure*}[t!]
\centering{\includegraphics[width = \textwidth, height=10.5cm]{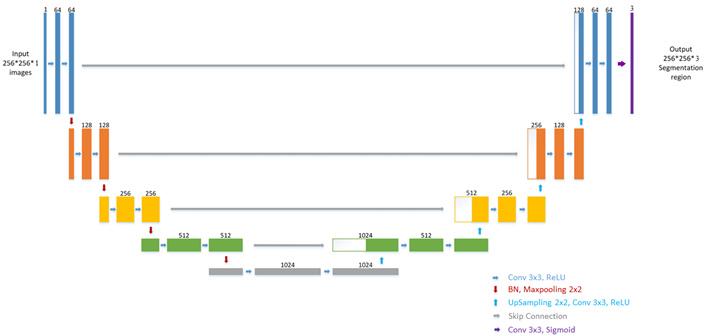}}
\caption{Multi Class U-Net Network Architecture}
\end{figure*}
\subsubsection{Multi Class U-Net Network Architecture}

U-Net model architecture is originally proposed by [Ronneberger et al., 2015] \cite{ronneberger2015u} for binary segmentation of gray-level images. It is composed of an encoding (contracting) path,  a bottleneck in the center, and an upsampling (expansive) path.
The architecture is shown in the Figure 5.
In this work, we used the multi-class U-Net model to jointly segment the optic cup and disc of a retinal image. 

The contracting part consists of four pairs of 3x3 convolutional layers with zero padding, where each pair convolutional layer is followed by a 2 × 2 max-pooling. Batch normalization [Ioffe and Szegedy, 2015] \cite{ioffe2015batch} and ReLU non-linearity is used in all layers to improve learning. The encoder path reduces the spatial dimensions (resolution) in every layer and doubles the number of channels (feature map) in every layer. 
The input resolution of the first layer of the network is set to  256x256x1 to match the resolution of input images.
The expansive path comprises four blocks of 2 × 2 transpose convolution of the feature map, which halves the number of feature channels and pairs of convolutional layers (3 × 3 convolutions with zero padding). Each pair of a convolutional layer is followed by ReLU non-linearity unit and batch normalization.
Additionally, corresponding pairs of convolutional layers in the contracting and expansive parts are connected by skip-connections. 
The skip-connections are simple concatenations of feature maps that help successive convolution layers in the up-sampling part to learn to produce better localized output, allowing the network to perform more precise segmentation. 
In the last layer of the proposed network, Softmax is used to select the best scoring category. To enable multi-class segmentation, the output segmentation layer is expanded from 1 to N feature maps, where N is the number of classes.

\subsubsection{Loss Function}
Since we implemented multi-category segmentation in our proposed model, we used $keras.utils.to$\textunderscore$categorical$ to process the data into categories. In our work, we consider the output into 3 categories. 
The training of the proposed network is accomplished in a fully supervised manner by minimizing the standard categorical cross-entropy function on a pixel-wise basis: 
\[L= -\frac{1}{N}\sum_{i=1}^{N}[yi \log(\hat{yi})+(1 -yi)\log(1 -\hat{yi})]\],where $yi$ are true labels, $\hat{yi}$ predicted labels, and $N$ is the number of classes.

\subsection{\textbf{Feature Extraction}}

A list of features are extracted and analysed from segmented binary images. These extracted features are then used for the classification of the retinal fundus images into Glaucomatous and Non-Glaucomatous.

\subsubsection{Cup-to-Disc Ratio Calculation}
The two main features that are primarily used for the classification of retinal fundus images for detection of Glaucoma are \textbf{Area Cup-to-Disc Ratio ($ACDR$)} and \textbf{Diameter Cup-to-Disc Ratio($DCDR$)}. The $ACDR$ is calculated by dividing the area of the optic cup with the area of the optic disc. The optical disc and the cup area are measured by considering the largest possible area inside the contours of the disc and cup, which are detected from their segmented images. Similarly, the $DCDR$ is determined by using the length of the major axis of the optic cup divided by the length of the major axis of the optic disc. The length of the major axis of the optic cup is $Cup Diameter$ while the length of the major axis of the optic disc is $Disc Diameter$. The results of $ACDR$, $Disc Diameter$, $Cup Diameter$, $Disc Area$, $Cup Area$ and $DCDR$ are used as parameters for detecting Glaucoma.

\subsubsection{Notching Feature Extraction}
A notch in the optic cup is another important factor that differentiates normal from glaucomatous eyes. Notching is defined as a phenomenon that causes a focal enlargement of the cup.
The most popular approach to detect notching is to apply the ISNT rule  \citep{harizman2006isnt}, which states that in a normal eye, the rim width varies from thickest to thinnest in the order: inferior (I), superior (S), nasal (N), and temporal (T).  In the cases of glaucomatous eyes, the ISNT rule is violated as the inferior rim appears to be thinner than normal. In general, in the case of Glaucoma affected eyes, the inferior region is affected first, followed by the superior, then the temporal, and at last nasal region. 
Since inferior and superior tissues are changed first, notching is mostly found in one of these quadrants, which often causes the optic cup to enlarge vertically or oblique to the optic nerve head.  
\\In this work, the notching feature extraction is done by calculating rim thickness profile $T\theta$ for the range [0, \ang{360}] in increments of \ang{1} as:
\[{T\theta}= d(D\theta, a) - d(C\theta, a)\],where \textbf{\textit{d(m,n)}} is the Euclidean distance between $m$ and $n$, \textbf{\textit{a}} is the Optic Disc centre, \textbf{\textit{D}}\textbf{$\theta$} is the point on the Optic Disc boundary which is at an angle $\theta$ from $a$ and \textbf{\textit{C}}\textbf{$\theta$} is the point on the Optic Cup boundary which is at an angle $\theta$ from $a$.

\begin{figure}[h]
\centering\includegraphics[width=0.5\linewidth]{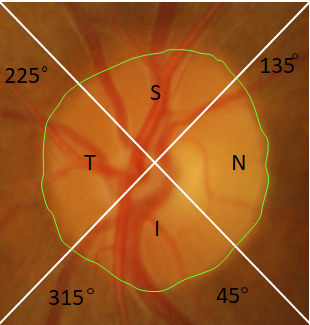}
\caption{The quadrant division of ISNT for a left eye.}
\end{figure}
To make $T\theta$ scale-invariant, $T\theta$ is divided by the length of the major axis of the segmented Optic Disc $k$ to get $X\theta$.
\[{X\theta} = \frac{T\theta}{k}\]
$X\theta$ is divided into four quadrants as depicted in Figure 6. The quadrant with angle $\theta$ in the range [\ang{1}, \ang{45}] and [\ang{316}, \ang{360}] forms the superior quadrant, the one with angle in the range [\ang{136}, \ang{225}] forms the inferior quadrant, and the angle in the range [\ang{46}, \ang{135}] forms temporal quadrant.
Finally, to capture the overall decrease in rim thickness in the Inferior and Superior quadrants, the mean of $X\theta$ in these two quadrants is computed separately, and the results are stored as I-distance and S-distance.

\subsection{\textbf{Decision Making with Machine-Learning Algorithms}}

The features extracted from the segmented images are used to train the Classification Model. In this work, we have used a supervised learning model, the support vector machine (SVM) classifier \citep{hearst1998support} for distinguishing normal eye fundus from glaucoma-affected eye fundus. The goal of the SVM algorithm is to build the best line or decision boundary that can divide n-dimensional space into classes so that we can easily assign a new data point in the correct category. SVM algorithm usually has a fast prediction speed even though it takes a long time to train from a training data set and reasonably high memory usage.
Since, in this case, the training dataset is small, SVM takes less time to train from the training dataset along with high prediction accuracy.
SVM has different kinds of kernel functions that are used to transform non-linear  to  linear separating  hyperplane  in  higher  dimensional
feature space.
In this paper, we have used RBF kernel as a kernel function.
\textbf{RBF kernel} is defined as:
\[{k(m,n)}= exp(-\gamma(||m-n||^2))\],where k is the function used for the transformation,
\\$\gamma$ is the free parameter and $\gamma$ = $\frac{1}{2\sigma^2}$,
\\$||m-n||$ gives the Euclidean distance between the
two features vectors.

This algorithm classifies the images in the database based on eight extracted features which are - \textbf{Area CDR, Diameter CDR, Cup Diameter, S-Distance, I-Distance, Disc Diameter, Cup Area, and Disc Area }. 
\section{Experimental Evaluation}

\subsection{Implementation details}
All experiments are executed in Python.  The network was realized in TensorFlow using the Keras wrapper. The network was trained for 100 epochs with a batch size of 2, using the Adam \citep{kingma2014adam} optimizer with default parameters. The validation set was used to evaluate the network and to prevent overfitting. Both training and validation images were resized to the resolution of 256×256 pixels, to match the input resolution of the network.

\subsection{Evaluation Metrics}
The segmentation results are verified with the ground truths of the experts provided in the dataset. 
In our paper, we adopt the Accuracy, Jaccard, F1-score, Recall, and Precision to evaluate the segmentation performance of the proposed technique.
The evaluation metrics are defined mathematically as:` 
\[Precision \; (PREC.)=\frac{TP}{TP + FP}\]
\[Recall \;=\frac{TP}{TP + FN}\]
\[F1_{score}=\;2\;\frac{Precision\;.\;Recall}{Precision \; + \; Recall}\]
\[Jaccard \;=\frac{TP}{TP + FP + FN}\]
\[Accuracy \;(ACC.) =\frac{TP + TN}{TP + TN + FP + FN}\]\\

where $TP$, $FP$, $TN$, $FN$ denotes True Positive, False Positive, True Negative, False Negative respectively.\\

The prediction accuracy of classification method has been represented by confusion matrix where horizontal axis represents True Condition and vertical axis represents Predicted Condition as shown in Table-1.\\
For Classification, we have used Precision, Specificity, Sensitivity, Accuracy, and Negative Predictive Value (NPV) for validating the results of glaucoma detection. The mathematical expressions of the parameters are denoted as:
\[Negative \; Predictive \; Value\;(NPV.)=\frac{TN}{TN + FN}\]
\[Specificity \; (SPEC.)=\frac{TN}{TN+FP}\]
\[Sensitivity \; (SENS.)=\frac{TP}{TP+FN}\]\\

where $TP$, $FP$, $TN$, $FN$ denotes True Positive, False Positive, True Negative, False Negative respectively.\\

\subsection{Segmentation Results}

To verify the effectiveness of our proposed technique, we evaluated our network on testing images of DRISHTI-GS dataset. The final scores of the evaluation metrics are the average of all testing images of the dataset.
Our method has achieved Accuracy, F1-score, Jaccard, Recall, and Precision of 0.99513, 0.89728, 0.82051, 0.91459, and 0.89498 on Optic Cup(OC) segmentation and 0.99697, 0.95229, 0.90973, 0.98103, and 0.92654 on Optic Disc(OD) segmentation.
Four retinal fundus images are selected from the test dataset arbitrarily and the visual representation of segmentation results of optic disc and cup regions are shown in Figure 7.

\begin{figure}[h]
\centering\includegraphics[width=1\linewidth, height=5cm]{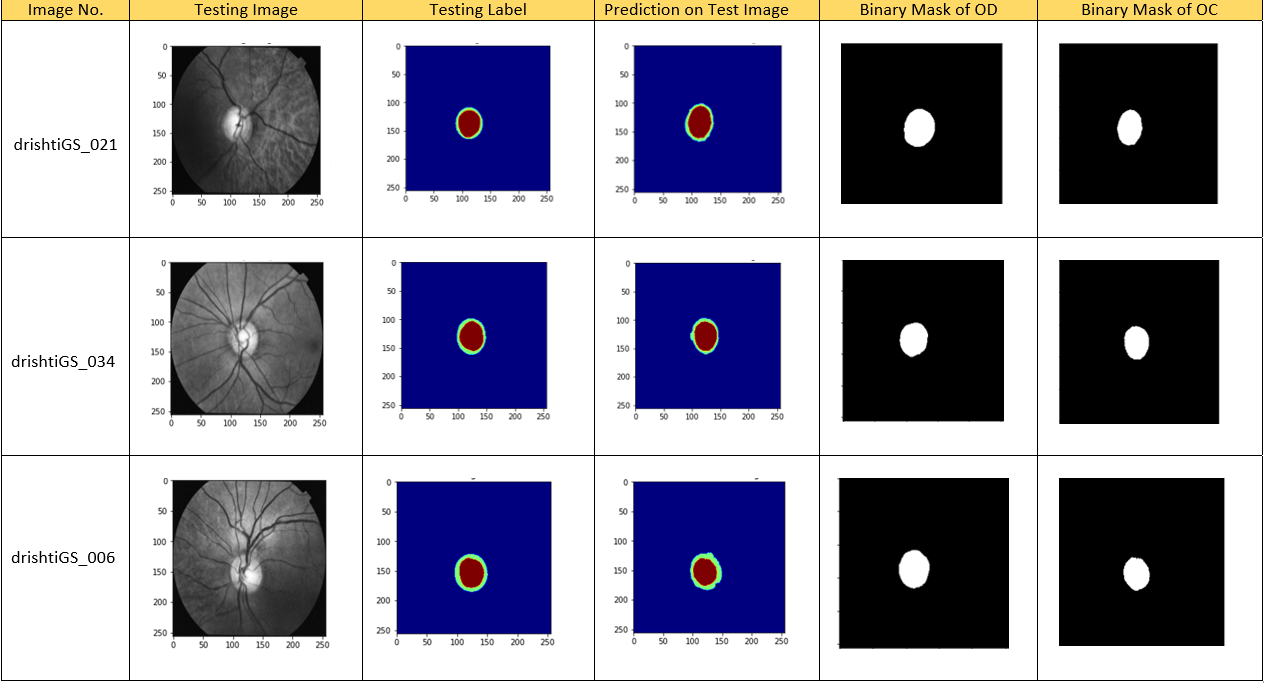}
\caption{First Column: Image No.; Second Column: Test Image in Grayscale; Third Column: Ground Truth of Image; Fourth Column: Prediction on Test Image; Fifth Column: Binary Mask of Segmented OD; Sixth Column: Binary Mask of Segmented OC)}
\end{figure}

\subsection{Classification Results}

The Support vector based supervised machine learning algorithm is used for the decision-making in glaucoma classification. Since SVM has high computational efficiency and performs well for a small training dataset, SVM has performed better in the classification of glaucoma and normal cases than other supervised learning methods.\\
\begin{table}[!h]
\centering
  \begin{adjustbox}{max width=0.48\textwidth}
\begin{tabular}{ll|c|c|c}
\cline{3-4}
                                                                                                              &                                                                                               & \multicolumn{2}{c|}{\textbf{True Condition}}                                                                      & \multicolumn{1}{l}{}                                                         \\ \cline{2-4}
\multicolumn{1}{l|}{}                                                                                         & \textbf{Total Population}                                                                     & \textbf{Glaucoma}                                        & \textbf{Normal}                                        & \multicolumn{1}{l}{}                                                         \\ \hline
\multicolumn{1}{|l|}{\multirow{2}{*}{\textbf{\begin{tabular}[c]{@{}l@{}}Predicted\\ Condition\end{tabular}}}} & \multicolumn{1}{c|}{\begin{tabular}[c]{@{}c@{}}Predicted\\ Condition\\ Glaucoma\end{tabular}} & \begin{tabular}[c]{@{}c@{}}TP\\ 10\\ \end{tabular} & \begin{tabular}[c]{@{}c@{}}FP\\ 1\\\end{tabular}  & \multicolumn{1}{c|}{\begin{tabular}[c]{@{}c@{}}PREC.\\ 90.9\%\end{tabular}} \\ \cline{2-5} 
\multicolumn{1}{|l|}{}                                                                                        & \multicolumn{1}{c|}{\begin{tabular}[c]{@{}c@{}}Predicted\\ Condition\\ Normal\end{tabular}}   & \begin{tabular}[c]{@{}c@{}}FN\\ 1\\ \end{tabular} & \begin{tabular}[c]{@{}c@{}}TN\\ 18\\\end{tabular} & \multicolumn{1}{c|}{\begin{tabular}[c]{@{}c@{}}NPV.\\ 94.73\%\end{tabular}}  \\ \hline
                                                                                                              &                                                                                               & \begin{tabular}[c]{@{}c@{}}SENS.\\ 90.9\%\end{tabular}  & \begin{tabular}[c]{@{}c@{}}SPEC.\\ 94.73\%\end{tabular}   & \multicolumn{1}{c|}{\begin{tabular}[c]{@{}c@{}}ACC.\\ 93.33\%\end{tabular}} \\ \cline{3-5} 
\end{tabular}
\end{adjustbox}
\renewcommand{\thetable}{1}
\caption{Confusion Matrix Representation for classification of Digital Fundus Image with Notching and CDR Features}
\end{table}
 Table. 1 depicts the confusion matrix where the dataset is classified into glaucoma and non-glaucoma cases using eight features: Area CDR, Diameter CDR, Cup Diameter, S-Distance, I-Distance, Disc Diameter, Cup Area, and Disc Area. Our classification method achieved a Sensitivity of 90.9\% and Specificity of 94.73\%, with an Accuracy of 93.33\%. 

\section{Conclusion}
In this paper, we presented a deep learning framework for joint segmentation of OD and OC regions from retinal fundus images, thereby extracting distinctive features necessary for decision-making in detecting glaucoma. The proposed algorithm employs a multi-class semantic segmentation model with UNet as its backbone network for segmentation and support vector based supervised learning model (SVM) for classification. CDR along with notching features of DFIs are used as primary parameters for classification of glaucoma. The proposed pipeline was implemented and tested using DRISHTI-GS dataset. The experimental results indicate that our network can improve the accuracy of Computer-aided design (CAD) systems in analyzing glaucoma, which can assist clinicians in the early detection and diagnosis of glaucoma. 
In future research work, we intend to apply our method to other medical image analysis tasks like brain tumor detection, lung cancer detection, etc. Besides this, we will also try to implement our method using a dataset that includes clinical information of patients like age, sex, race to improve the generalization performance of the proposed algorithm.\\

\bibliographystyle{ieeetr}
\bibliography{main.bib}
\end{document}